\documentstyle[11pt]{article}
\begin{document}
\centerline{\Large\bf A Simple Proof of the Unconditional Security}
\vskip 2mm
\centerline{\Large\bf of Relativistic Quantum Cryptography}
\vskip 3mm
\centerline{S.N.Molotkov and S.S.Nazin}
\centerline{\sl\small Institute of Solid State of Russian Academy of Sciences}
\centerline{\sl\small Chernogolovka, Moscow District, 142432, Russia}
\vskip 3mm

\begin{abstract}
A simple proof of the unconditional security of a relativistic quantum
cryptosystem based on orthogonal states is proposed. Restrictions imposed by
special relativity allow to substantially simplify the proof compared with
the non-relativistic cryptosystems involving non-orthogonal states.
Important for the proposed protocol is the spatio-temporal structure of the
quantum states which is actually ignored in the non-relativistic protocols
employing only the structure of the state space of the information carriers.
As a consequence, the simplification arises because of the inefficiency of
collective measurements which constitute the major problem in the 
non-relativistic case.
\end{abstract}
\vskip 3mm

\section*{Introduction.}
In contrast to the classical cryptography based on the laws of classical
physics, quantum cryptography employs the fundamental laws of non-relativistic 
quantum mechanics [1,2]. Security of classical cryptography is 
based on the unproved computational complexity of some 
functions, for example the discrete logarithm, which are 
assumed to be computed using a physical device operating 
according to the laws of classical physics. It is also 
implicitly assumed that the exchange of information between the 
legitimate users is performed employing the classical objects.
Since the laws of classical physics do not prohibit 
simultaneous measurements of any dynamical variables of a 
classical system without any disturbances, it is impossible to 
reliably detect the eavesdropping attempts during the exchange 
of information between the legitimate users within the 
framework of classical physics. Therefore, the security of 
quantum cryptosystems cannot be based on the detection of 
eavesdropping attempts during the key distribution and can only
be associated with the exponential computational complexity
of some functions.

The laws of classical physics constitute only an approximate 
description of physical reality. A more accurate description is 
provided by the non-relativistic quantum mechanics. The quantum 
mechanics allow in principle to construct a computational 
device (quantum computer) which for some problems is more 
powerful than the classical computers. The factorization 
problem which is faced by eavesdropper when attempting to 
extract (compute) the key in some open key classical 
cryptosystems is unsolvable (possesses exponential complexity) 
for a classical computer, but becomes solvable (possesses
only polynomial complexity) for a quantum computer [3].
Therefore, the laws of non-relativistic quantum mechanics do 
not allow the unconditionally secure information exchange using 
the available classical algorithms (the unconditional security 
is understood as the security following from the fundamental 
laws of nature rather than that based on computational 
complexity).

Prohibiting the unconditionally secure (in the above sense) classical 
cryptography, the quantum mechanics opens new possibilities for quantum
cryptography which is based on reliable detection of the eavesdropping 
attempts guaranteed by the laws of quantum mechanics if the information 
is carried by quantum systems.

The non-relativistic quantum cryptography is based on the following two
consequences of the postulates of non-relativistic quantum mechanics:
\begin{itemize}
\item{} The unknown quantum state cannot be copied (no cloning theorem) [4].
\item{} No information about the states belonging to a non-orthogonal
        basis can be obtained without disturbing them [5].
\end{itemize}

For the orthogonal states this restriction does not hold.
Moreover, there exist no restrictions on the instantaneous distinguishability
of the orthogonal states without any disturbance. Therefore the
usage of orthogonal states in the non-relativistic quantum cryptosystems
has not even been discussed.

The non-relativistic quantum cryptosystems actually do not employ
the spatio-temporal structure of the quantum states (since both the
no-cloning theorem and the statement concerning the impossibility of 
acquiring any information about the states belonging to a non-orthogonal 
basis without disturbing them have a completely general nature).
Used in the non-relativistic quantum protocols are only the
properties of the Hilbert state space of the information carriers.
The effects of state propagation between two distant users are not
taken into account explicitly since they are not important.
To be more precise, the attempts to take into account the spatio-temporal
structure of the states do not seem promising from the cryptographic point 
of view because of the absence of any restrictions on the maximum speed of
information transfer in the non-relativistic theory.

A rigorous proof of the unconditional security of non-relativistic quantum
protocols when both the eavesdropper and legitimate users are only limited 
by the laws of quantum mechanics presents a very difficult problem.
Several proofs characterized by different levels of complexity and 
employing different assumptions have been proposed until now. However,
because of the complexity of the problem, no generally accepted
view on the proof of the unconditional security has yet evolved [6,7,8,9].

Just as the classical physics, the non-relativistic quantum mechanics 
provides only an approximate description of nature. A more accurate 
description taking into account the spatio-temporal structure of quantum 
states and the restrictions imposed by general relativity is provided by
the relativistic quantum theory which, because of the absence of any 
sensible interpretation of relativistic quantum mechanics, 
arises from the very outset as the quantum field theory.

The quantum field theory does not close the non-relativistic quantum 
cryptography because the states of quantum field systems are also described
by the rays in physical Hilbert space [10], just as in the non-relativistic 
quantum mechanics. Since the non-relativistic quantum cryptographic protocols
employ only the properties of the states in the Hilbert space, the 
non-relativistic protocols survive in the quantum field theory.
For the cryptographic problems the quantum field theory can only 
contribute a substantially new aspect if the spatio-temporal structure
of the states is explicitly considered in the protocols. To be more precise,
one should take into account that the quantum states, although being described
by the rays in the Hilbert space just as in the non-relativistic quantum 
mechanics, are generated by the field operators (to be more precise,
operator-valued distributions) which contain information about the
structure of the space-time. The field operators satisfy the commutation 
relations expressing the microcausality principle. The latter follows from the 
restrictions imposed by special relativity according to which there should be
no causal relation between any two points of Minkowski space-time separated 
by a space-like interval. In addition the field theory allows to explicitly
take into account and employ in the construction of cryptographic protocols
the effects of state propagation in the Minkowski space-time.

The restrictions imposed by the filed theory and special relativity
result in substantial differences in some problems of quantum information
theory (see details in Refs. [11,12]) compared with the non-relativistic case.

Presented below is a simple example of an unconditionally secure
relativistic quantum cryptosystem in a noisy channel employing the
quantum field states (photons) as the information carriers. Because of
more restrictive laws of the relativistic quantum field theory,
the proof of the unconditional security is substantially simplified
compared with the non-relativistic case. In addition, the proposed
scheme can be rather easily realized experimentally, in contrast to
the unconditionally secure exchange protocols in the non-relativistic
case where the proof of security substantially employs the collective
measurements which are very difficult for experimental realization.
The proposed scheme explicitly employs the causality effects and only
the individual measurements.

The below arguments are induced by the paper [13] which, in our opinion,
has not been assessed correctly [14]. Further improvements [15] have in fact 
reduced it to a non-relativistic cryptosystem based on non-orthogonal
states actually discarding the key new idea.

The key idea is to use the internal degrees of freedom of a quantized field
(the photon helicity) to code the transmitted information while the 
spatial degrees of freedom are used to detect the eavesdropping attempts.
This is a substantially new aspect compared with the non-relativistic
case which, combined with the special relativity, allows to reliably detect
any attempts by eavesdropper to delay the transmitted states and to guarantee
that the eavesdropper obtains no information about them. The fact that the 
transmitted states are the states of a quantum field is also important for 
the protocol.

The quantum field theory allows to use even orthogonal states in the
cryptosystems. Since in the relativistic case the system states are
also described by the rays in the Hilbert state space just as in the 
non-relativistic quantum mechanics, both the no-cloning theorem
and the statement on impossibility of reliable distinguishability of 
non-orthogonal states without disturbing them remain valid because
their proofs employ only the properties of the Hilbert state space.

Important for further analysis are the following two circumstances 
dictated by the quantum field theory (see details in Refs. [16,17]):
\begin{itemize}
\item{} Reliable distinguishability of a pair of orthogonal states
of a free quantum field requires access to the entire domain of the 
Minkowski space-time where the state supports are located. Formally, any two 
orthogonal states of a free quantum field can only be reliably distinguished 
without disturbing them in an infinite time because of the inherent 
unlocalizability of quantum field states (no field state with a compact support
in the position space can be constructed whose support in the four-dimensional 
momentum space is defined on the mass shell).
\item{} The theory allows the existence of free quantum field states
arbitrarily strongly localized in the space-time (with any localization 
degree and decaying at infinity according to a law arbitrarily close to 
the exponential one). 
\end{itemize} 
The fundamental unlocalizability of the states is a consequence of the
local nature of quantum field theory.

The latter circumstance is important for cryptography since it allows 
to prepare arbitrarily strongly localized states which have any required 
arbitrarily small tails decaying arbitrarily close to the exponential law
beyond the space-time domains controlled by the legitimate users.
The probability of distinguishing two orthogonal states of a quantum field 
can vary due to the effects of field propagation from the controlled domain
of the Minkowski space-time to the domain accessible for the measurement 
procedure from 1/2 (complete indistinguishability) to 1 (reliable 
distinguishability). To be more precise, due to the preparation of
strongly localized field states, the probabilities of obtaining
the measurement results in a finite time can be made arbitrarily close 
to 1/2 or 1, the deviation being the smallest parameter in the considered
problem. Because of the restriction on the maximum speed of both field 
propagation and motion of classical objects (including measuring apparatus),
the access to the entire domain of the existence of the field consisting
of two arbitrarily strongly localized but spatially separated ``halves''
inevitably requires a finite time and cannot be performed instantly.
Therefore, as the information is being acquired, the distinguishability
probability changes from 1/2 to 1 in a finite time interval. Strong
localization of the field ``halves'' separated by $\tau_d$ allows to 
construct the protocol in such a way that the distinguishability
probability $P(\tau)=1/2$ at $0\le\tau< \tau_d$ and then jumps to 
$P(\tau)=1$  at $\tau=\tau_d$ (the jump is smeared by the state localization
length $\Delta\tau\ll\tau_d$). The jump smearing is controlled by
the localization of each ``halves'' and can be made arbitrarily small.

Since the relativistic quantum cryptography explicitly employs the
spatio-temporal structure of the states, the protocol security
cannot be proved without taking the geometry of a specific cryptosystem
into account, in contrast to the non-relativistic case where only
the structure of the state space is used.

\section*{Measurements used in the protocol.}
Consider now the measurements used in the protocol. Since we should 
necessarily take into account the specific geometry of the cryptosystem,
we shall consider a simple one-dimensional model containing all the important 
features dictated by the field theory. We shall consider the particles 
(field quanta) moving with the speed of light and the spectrum residing on 
the one-dimensional mass shell (front part of the one-dimensional light 
cone in the momentum representation, $k^2-k_0^2=0$).

Each of the users A and B controls the neighbourhoods of points $x_{A,B}$ 
(see the figure). The sizes of the controlled domains are dictated by the 
localization degree of the states and can be made arbitrarily small 
(should be of the order of the state localization length). In this geometry
it is sufficient to consider the states propagating from  $x_A$ to $x_B$
with positive momenta $k>0$. All the states are defined on one part 
of the light cone $\tau=x-t$ ($c=1$). 
The Hilbert state space of the information carriers is 
${\cal H}_k\otimes {\bf C}^2$, where  ${\cal H}_k$ is related to the
spatial while ${\bf C}^2$ to internal (polarizational) degrees of freedom. 

Consider a pair of orthogonal states 
\begin{equation}
|\psi_{0,1}\rangle=\frac{1}{\sqrt{2}}\int_{0}^{\infty} 
f(k)|k\rangle\otimes(|+\rangle \pm |-\rangle) dk,
\end{equation}
where $|\pm\rangle\in{\bf C}^2$ are the orthogonal basis states, and
\begin{displaymath}
\langle k|k'\rangle=\delta(k-k'),\quad k,k'>0,\quad
\int_{0}^{\infty}|f(k)|^2 dk=1.
\end{displaymath}
In the position representation the states on the branch of the 
one-dimensional light cone have the form
\begin{equation}
|\psi_{0,1}\rangle=\frac{1}{\sqrt{2}}\int_{-\infty}^{\infty} 
f(\tau)|\tau\rangle\otimes(|+\rangle \pm |-\rangle) d\tau,
\quad\tau=x-t,
\end{equation}
\begin{displaymath}
|\tau\rangle=\int_{0}^{\infty}\mbox{e}^{-ik\tau}|k\rangle dk,
\quad
f(\tau)=\int_{0}^{\infty}\mbox{e}^{ik\tau} f(k)dk.
\end{displaymath}
The basis $\{|\tau\rangle\}$ is not orthogonal, 
$\langle\tau|\tau'\rangle=\delta_+(\tau-\tau')\neq \delta(\tau-\tau')$.

The measurement allowing to reliably distinguish the specified pair
of orthogonal states is given by an appropriate identity resolution in 
${\cal H}_k\otimes {\bf C}^2$ (and, similarly, by a resolution
in ${\cal H}_{\tau}\otimes {\bf C}^2$
where ${\cal H}_{\tau}$ is the space isomorphic to ${\cal H}_k$ 
spanned by the basis $\{|\tau\rangle\}$). 

The measurement is given by
\begin{equation}
{\cal M}_0 + {\cal M}_1=I_k\otimes I_{{\bf C}^2},
\quad
{\cal M}_{0,1}=I_k\otimes {\cal P}_{0,1}=I_{\tau}\otimes {\cal P}_{0,1},
\end{equation}
\begin{displaymath}
{\cal P}_{0}=|0\rangle\langle 0|,\quad 
{\cal P}_{1}=|1\rangle\langle 1|,\quad 
|0\rangle=\frac{1}{\sqrt{2}}\left(|+\rangle + |-\rangle\right),\quad
|1\rangle=\frac{1}{\sqrt{2}}\left(|+\rangle - |-\rangle\right),
\end{displaymath}
\begin{equation}
I_k=\int_{0}^{\infty}|k\rangle\langle k|dk=I_{\tau}=
\int_{-\infty}^{\infty}|\tau\rangle\langle\tau|d\tau=\int_{-\infty}^{\infty}
{\cal M}(d\tau),
\end{equation}
\begin{displaymath}
{\cal M}(d\tau)=
\left( \int_{0}^{\infty}\mbox{e}^{-ik\tau}|k\rangle dk \right)
\left( \int_{0}^{\infty}\mbox{e}^{ik'\tau}\langle k'| dk' \right) d\tau.
\end{displaymath}
Note that the measurement (4) is non-local on the light cone.

Support of the state $f(\tau)$ on the light cone can be chosen arbitrarily 
strongly localized, in the extreme case $|f(\tau)|^2\rightarrow \delta(\tau)$. 
Strictly speaking, in the field theory the states defined on the mass shell
cannot have a compact support in the Minkowski space-time. However,
one can construct arbitrarily strongly localized states with the tails 
arbitrarily close to the exponential ones [17]. The latter means that it is
possible to choose a time window  $\Delta \tau$ such that the probability of
detecting the state in this window can be made arbitrarily close to unit.
We shall assume that the state (actually the packet shape $f(\tau)$) and
the time interval $\Delta\tau$ are chosen such that the probability of
detecting the state outside the time window $\Delta\tau$ due to the state
tails extending beyond $\Delta \tau$ can be made exponentially arbitrarily 
close to zero. In the rest of the paper we shall assume this parameter
to be the smallest one in the problem. To be more precise, the probability of
detecting the input states $|\psi_{0,1}\rangle$ in the time window $\Delta\tau$
in the channel 0 (the channel corresponding to ${\cal P}_0$) and, 
respectively, in the channel 1 (${\cal P}_1$) (see Eq.(4)) is 
\begin{displaymath}
\mbox{Pr}\{\Delta\tau, |\psi_{0,1}\rangle\}=\mbox{Tr}
\left\{
\left(\left( \int_{-\Delta\tau}^{\Delta\tau}{\cal M}(d\tau) \right)
\otimes
{\cal P}_{0,1}\right)
|\psi_{0,1}\rangle\langle\psi_{0,1}|
\right\}=
\int_{-\Delta\tau}^{\Delta\tau} |f(\tau)|^2d\tau=1-\delta,
\end{displaymath}
where $\delta$ is only different from zero due to the state tails 
\begin{displaymath}
\int_{|\tau|>\Delta\tau} |f(\tau)|^2d\tau=\delta\rightarrow 0.
\end{displaymath}

In other words, there are two different parameters in the problem:
$\Delta\tau$, the typical state localization (the interval is chosen 
so that the squared amplitude state $f(\tau)$ integrated over it is 
arbitrarily close to unit) and $\tau_d$, the separation between the two 
``halves'' of the state chosen to make the overlap between the tails of the two 
``halves'' arbitrarily small ($\Delta\tau\ll \tau_d$).
In the further analysis we shall assume for convenience
(with the above reservations) the support to be a compact set
since this assumption will not affect the final result.

It should be noted here that {\it the measurement under consideration 
cannot be interpreted as the one having duration $\Delta\tau$.} 
In each measurement act the obtained result (reading of a classical device,
e.g. of a fast photodetector with a small response time (in the limit, 
formally infinitely small) operating in the waiting mode) arises at a random
time moment in the interval $(\tau,\tau+d\tau)$ with the probability density
\begin{displaymath}
\mbox{Pr}\{d\tau, |\psi_{0,1}\rangle\}=\mbox{Tr}
\left\{
\left(\left( {\cal M}(d\tau) \right)
\otimes
{\cal P}_{0,1}\right)
|\psi_{0,1}\rangle\langle\psi_{0,1}|
\right\}=|f(\tau)|^2d\tau.
\end{displaymath}
This interpretation is very natural and is consistent with the classical 
limit when a classical signal with the temporal shape $f(\tau)$ is
measured.

Let the states $|\psi_0\rangle$ (or $|\psi_1\rangle$) are prepared  
at time $\tau_i$ (to within $\Delta\tau$; this requires the control 
over the spatio-temporal domain of size $\Delta\tau$)
\begin{equation}
|\psi_{0,1}\rangle=
\frac{1}{\sqrt{2}}\int_{-\infty}^{\infty}
(f(\tau-\tau_i)|\tau\rangle
\otimes \left( |+\rangle\pm |-\rangle)\right) d\tau,
\end{equation}
and then a unitary transformation is performed which does not depend
on the prepared state:
\begin{equation}
|\psi_{0,1}(\tau_d)\rangle=
U|\psi_{0,1}\rangle=
\frac{1}{\sqrt{2}}\int_{-\infty}^{\infty}
\left(
f(\tau-\tau_i-\tau_d)|\tau\rangle \otimes |+\rangle \pm 
f(\tau-\tau_i)|\tau\rangle\otimes |-\rangle
\right)
d\tau.
\end{equation}
The accuracy of the moment of preparation is determined by
the width of the state support.

Matrix elements of the unitary operator have the form
\begin{equation}
\langle +|\langle \tau'|U|\tau\rangle|+\rangle=
\delta_+(\tau-\tau'-\tau_d)=\int_{0}^{\infty}\mbox{e}^{ik(\tau-\tau'-\tau_d)} dk,
\end{equation}
\begin{displaymath}
\langle -|\langle \tau'|U|\tau\rangle|-\rangle=\delta_+(\tau-\tau'),
\quad
\langle \pm|\langle \tau'|U|\tau\rangle|\mp\rangle=0.
\end{displaymath}
This unitary transformation is non-local on the light cone $\tau=x-t$ (for 
$\tau_d\neq 0$) which means that its realization requires access to a
domain of the size $\tau_d$ (to within $\Delta\tau$) on the light cone.
Physically, this unitary transformation corresponds to a shift
(delay) of the ``half'' of the state with the $|-\rangle$ polarization
along the light cone. If this transformation is performed at fixed $x$
(locally in the ordinary position space), then it requires time
$\Delta t=\tau_d$ (since $\tau=x-t$), or it requires the spatial domain of 
size $\Delta x=c\tau_d$ ($c=1$) if the transformation is performed
at a fixed moment of time but non-locally in space (in a spatial domain of 
size $\Delta x=c\tau_d$). 

The spatio-temporal interval $\tau_d$ on the branch of the light cone
does not depend on the choice of the reference frame since the light cone
is Lorentz-invariant. Hence the eavesdropper cannot use the
twin paradox [16].

The extended states $|\psi_{0}(\tau_d)\rangle$ and $|\psi_{1}(\tau_d)\rangle$ 
are orthogonal.However, the orthogonality is a non-local property 
in the sense that to find out whether or not the two states are orthogonal
one has to have access to the spatio-temporal domain 
(interval) of length $\ge\tau_d$ to within $\Delta\tau\rightarrow 0$.
In other words, the orthogonality is also a non-local property in the
Hilbert space ${\cal H}_{\tau}$ in the sense that one has to have access
to all states $|\tau\rangle$ (access to the spatio-temporal domain
$\tau\ge \tau_d$), on which the space ${\cal H}_{\tau}$ is spanned.

All the non-relativistic quantum cryptographic protocols assume that
the entire Hilbert state spaces are always accessible to both the legitimate 
users and eavesdropper. In the relativistic case the access can be changed by
the effects of the state propagation from the domain controlled by the
legitimate user to the domain accessible to the eavesdropper.
``Extension'' of the state and the limitation on the maximum speed of motion 
of classical objects and propagation of quantum states allows to construct 
a protocol in such a way that the state is never available to the eavesdropper 
as a whole. To be more precise, the attempts to get access to the entire
state requires access to the whole interval where the state is present.
However, because of the finite speed of light the attempts to get access
to a finite spatio-temporal domain result in inevitable delay in the state 
detection by a legitimate user. This is actually the reason why the collective
measurements which are efficient in the non-relativistic case (because
of the availability of the entire state space to all the participants)
and which present the major obstacle in the proof of the unconditional 
security, are insignificant here. One can only consider the individual 
measurements in each transmission since the presence of eavesdropper is 
detected by the delay of state detection by a legitimate user in each event.

If one has access to the interval $2T$ centered about $\tau_0$,
$T_0=(-T+\tau_0,\tau_0+T)$ and $2T<\tau_d+\Delta\tau$, then no measurement 
over the states $|\psi_{0}(\tau_d)\rangle$ and $|\psi_{1}(\tau_d)\rangle$ 
can distinguish between them (the states seem to be identical). 
The latter is formally expressed by the restriction of the density matrix
to the subspace ${\cal H}_{T_0}$ spanned by the states $|\tau\rangle$
whose support belongs to the interval $T_0$. For the density matrix we have
\begin{equation}
\rho_T=\mbox{Tr}_{\tau}\left\{\left(\left(\int_{T_0}{\cal M}(d\tau)\right)
\otimes I_{{\bf C}^2}\right)
|\psi_{0,1}(\tau_d)\rangle \langle\psi_{0,1}(\tau_d)|\right\}=
\end{equation}
\begin{displaymath}
\frac{1}{2}\int_{T_0}|f(\tau)|^2d\tau\otimes |+\rangle\langle +|+
\frac{1}{2}\int_{T_0}|f(\tau-\tau_d)|^2d\tau\otimes |-\rangle\langle -|.
\end{displaymath}
If the interval $T_0<\tau_d +\Delta\tau$ does not cover simultaneously
the supports of the states with different polarizations (see figure)
then only one of the functions $f(\tau)$, or $f(\tau-\tau_d)$ 
is different from zero. Thus, no measurement in the spatio-temporal domain
(interval $T_0<\tau_d+\Delta\tau$) can distinguish the orthogonal states.
The probability of distinguishing the states is 1/2 (equal to
the probability of simply guessing). Because of the limitation on the
maximum speed the access to the interval $\tau_d$ cannot
be obtained faster than the length of the interval itself.

\section*{Description of the protocol.}

Legitimate users A and B control the neighbourhoods of points $x_A$ and $x_B$, 
$x_A< x_B$ (see the figure). Their clocks are assumed to be synchronized. 
The size of the controlled neighbourhoods should be 
$\Delta x_{A,B}\sim \Delta\tau$. The state support width 
$\Delta\tau\rightarrow 0$ is assumed to be known and is the smallest parameter
in the problem. The communication channel length ($x_B-x_A=\tau_{ch}$) 
is also assumed to be known (however, the absolutely exact knowledge of
$\tau_{ch}$ is not required).

{\bf 1.} User A prepares with equal probabilities one of the states 
corresponding to 0 or 1 at a random time $\tau_i$ (to within $\Delta\tau$) 
\begin{equation}
|\psi_{0,1}\rangle=\frac{1}{\sqrt{2}}\int_{-\infty}^{\infty}
f(\tau-\tau_i-\tau_A)|\tau\rangle\otimes(|+\rangle\pm |-\rangle) d\tau,
\quad \tau_A=x_A.
\end{equation}
Formally, the integration over $d\tau$ in Eq. (9) is performed over the 
entire branch of the light cone, although actually only the basis vectors
$|\tau\rangle$ from the interval $\Delta\tau$ are involved in the state 
formtaion.

{\bf 2.} The ``half'' of the state (with the component $|+\rangle$) 
is sent into the communication channel while the second ``half''
with the component $|-\rangle$ is delayed which is described by the
unitary transformation  
$U_A(\tau_d)$ (this interpretation of the unitary transformation is natural
since $U_A(\tau_d)$ has matrix elements relating the states shifted
along the light cone only for the polarization component $|+\rangle$)
\begin{equation}
|\psi_{0,1}(\tau_d)\rangle=
U_A(\tau_d)|\psi_{0,1}\rangle=\frac{1}{\sqrt{2}}\int_{-\infty}^{\infty}
\left(
f(\tau-\tau_i-\tau_A-\tau_d)|+\rangle\pm f(\tau-\tau_i-\tau_A)|-\rangle
\right)
\otimes|\tau\rangle d\tau.
\end{equation}
The transformation is non-local, its realization in the neighbourhood of
point $x_A$ requires time $\tau_d$ and does not depend on whether the state
0 or 1 is considered.

{\bf 3.} State propagation from A to B is formally described by a unitary
translation $U(\tau_{ch})$ along the branch of the light cone by the interval
$\tau_{ch}$ ($\tau_{ch}=x_B-x_A$)
\begin{equation}
|\psi_{0,1}(\tau_{ch})\rangle=
U(\tau_{ch})|\psi_{0,1}(\tau_d)\rangle=
\end{equation}
\begin{displaymath}
\frac{1}{\sqrt{2}}\int_{-\infty}^{\infty}
\left(f(\tau-\tau_i-\tau_A-\tau_d-\tau_{ch})|+\rangle\pm 
f(\tau-\tau_i-\tau_A-\tau_{ch})|-\rangle\right) 
\otimes |\tau\rangle d\tau.
\end{displaymath}

{\bf 4.} User B performs a unitary transformation $U_B(-\tau_d)$ which 
does not depend on the input state which merges the ``halves'' of the states
back together (the shift of the ``half'' with the component $|+\rangle$ 
from and towards $|-\rangle$ can be preformed using the beam splitters, mirrors,
and delay line)
\begin{equation}
U_B(-\tau_d)|\psi_{0,1}(\tau_{ch})\rangle=
\frac{1}{\sqrt{2}}\int_{-\infty}^{\infty}
f(\tau-\tau_i-\tau_A-\tau_{ch})
(|+\rangle\pm |-\rangle)\otimes |\tau\rangle d\tau.
\end{equation}
Matrix elements of the operator $U_B(-\tau_d)$ are similar to those of
$U_A(\tau_d)$, but with $\tau_d$ changed to $-\tau_d$
\begin{displaymath}
\langle +|\langle \tau'|U_B(-\tau_d)|\tau\rangle|+\rangle=
\delta_+(\tau-\tau'+\tau_d),
\end{displaymath}
\begin{displaymath}
\langle -|\langle \tau'|U_B(-\tau_d)|\tau\rangle|-\rangle=\delta_+(\tau-\tau'),
\quad
\langle \pm|\langle \tau'|U_B(-\tau_d)|\tau\rangle|\mp\rangle=0.
\end{displaymath}

{\bf 5.} After the transformation $U_B(-\tau_d)$ the user B preforms the
measurement realizing the identity resolution given by Eqs.(3,4). 
The space of all possible measurement outcomes (results) is the set 
$\Theta=\{i,\tau: i=0,1;\tau\in(-\infty,\infty)\}$ 
(index $i=0,1$ describes the outcomes in channels 0 and 1), 
\begin{equation}
\int_{-\infty}^{\infty}{\cal M}(d\tau)\otimes
\left( 
{\cal P}_0 +
{\cal P}_1 
\right)=
I_{\tau}\otimes I_{{\bf C}^2}.  
\end{equation}
This measurement describes the probabilities of obtaining
a result in the interval $\Delta\tau$ in the channel 0 or 1 
which is given by the expression
\begin{equation}
\mbox{Pr}\{(\Delta \tau)\}=\int_{\Delta\tau}
|f(\tau-\tau_i-\tau_A-\tau_{ch})|^2d\tau.
\end{equation}
The result is different from zero if the interval $\Delta\tau$ 
covers the state support. Because of the orthogonality of the states
the outcomes in the detection channels of user B coincide identically
with the values sent by user A (if the noise is neglected).

The choice $\tau_d>\tau_{ch}$ means that only a part of the state is always
available in the channel (only a part of the state space ${\cal H}_{\tau}$
is accessible), the access to only a part of the state space guaranteeing 
that the information about the states being zero (the probability of 
distinguishing the states is 1/2). Because of the restriction on the
maximum propagation speed the access to the second ``half'' of the state
requires access to the interval $\tau>\tau_d>\tau_{ch}$ which inevitably
results in a delay of the detection moment by the legitimate user B.

{\bf 6.} User B announces through the public channel the detection time
$\tau_B$ of the state (which is known to within $\Delta\tau \rightarrow 0$
because of the narrow support of $f(\tau)$). If the detection did not occur,
the transmission is discarded. After user B announced the detection 
time, user A announces the sending time $\tau_i$. If the detection time obtained 
by user B $\tau_B=\tau_i+\tau_d+\tau_{ch}$ (to within $\Delta\tau$), 
the attempt is accepted. If the detection time is found to by delayed
(by more than $\ge\tau_d$), the attempt is discarded. 

{\it It is fundamentally important, that the absence of delay in the detection 
moment measured by user B already guarantees that the eavesdropper has zero 
information about the states sent by user A (the probability of 
distinguishing the states is 1/2).} Deviation of this probability from 1/2
is due to the exponential tails of the states which can be made arbitrarily
small by separating the ``halves'' (increasing $\tau_d$). If the state
$f(\tau)$ has a compact support, the absence of a delay means that the 
information obtained by eavesdropper is strictly zero.

{\bf 7.} In the accepted transmissions the eavesdropper has zero 
information. However, because of the noise in the channel (decoherence 
processes) the sequences of 0 and 1 possessed by legitimate are not yet 
identical. The differences can be caused both by the eavesdropper
and by the natural noise. For example, the eavesdropper can detect
the first ``half'' of the state which is being sent by user A.
This detection requires time $\Delta\tau\rightarrow 0$
(hence the eavesdropper determines time $\tau_i$). 
Then immediately after the detection the eavesdropper arbitrarily prepares
his state $f(\tau)$ which will not cause any delay for user B but
result in the inconsistency with the state sent by user A that will be
interpreted as a noise in the channel.  It follows from Eq.(8)
that the probability of determination of the state preparation time
is 1/2 (because of the accessibility of only a ``half'' of the state),
the state detection itself giving to the eavesdropper zero information.
The probability of guessing the state is 1/2. The total probability of 
correct determination of the state in the channel is
$1/2\cdot 1/2=1/4$. It should be noted that the probability of correctly
guessing the state by the eavesdropper in each transmission when he
undertakes no attempt to obtain any information from the channel is 1/2.
At first glance this seems strange since the probability of correct 
determination of the state by the eavesdropper is lower by a factor of 2 
when he has access to the channel than when he simply guesses the state
without access to the communication channel. However, actually there is nothing
strange in this fact since the simple guessing involves only two outcomes 
(0 or 1). On the other hand, when detecting the state in the channel one 
should also determine the state preparation moment $\tau_i$ (so that this 
event is not discarded by the legitimate users) and the detection probability
for a ``half'' of the state is 1/2. Therefore, if the eavesdropper attempts
to access the channel the probability of erroneous interpretation of the 
intercepted state (which is equal 3/4) includes also the probability of 
erroneous determination of the very fact of sending any state at all 
in the considered time interval (which is equal 1/2): since the detection
itself yields zero information about the state, one has to guess it and the 
probability of guessing is 1/2.

{\it Hence, if the legitimate users keep only the transmissions
in which no delay in photon detection was observed (to be more prices,
the delay did not exceed $\tau_d+\tau_{ch}$) they can be sure that the 
probability of correct identification of the state by the eavesdropper
intruding into the communication channel does not exceed 1/4. The latter
is twice as low as the probability of correctly guessing the state
without intruding into the channel.}

We have here a curious situation which does not take place in the
non-relativistic case. If we remember that the aim of the eavesdropper
is to obtain maximum information about the key simultaneously minimizing
the probability of being detected by legitimate users, then in this sense
his optimal strategy consists in simply guessing the state sent in each
attempt (the access to the communication channel is not required).
It is only sufficient to have access to the classical communication channel
(in order to know the total number of accepted transmissions)
which in all quantum cryptographic problems is always assumed to be public.
In that case the probability of detecting the eavesdropper is of course zero
since he does not disturb the channel. Attempts to intrude into the
communication channel are only sensible if the probability of the state 
identification (restricted to the transmissions accepted by the 
legitimate users) exceeds the probability of simple guessing which is 1/2.

Hence the maximum probability of correct identification by the eavesdropper 
of each transmission accepted by the legitimate users is 1/2.

{\bf 8.} Now the only problem to be solved is to ensure that the legitimate 
users have identical keys. Consider first a noiseless channel. After
the sending session is completed and $2N$ transmissions
are accepted, the users A and B perform $m<2N$ rounds of random hashing
(parity check with a random bit string, see details in Ref. [18]).
The length of the original sequence of $2N$ bits is reduced by 2 after each 
round. If the parity check succeeds in $m$ rounds, the probability of having 
at least one difference in the left strings of $2N-2m$ bits possessed by users 
A and B does not exceed $2^{-m}$ which means that users A and B have identical 
keys with the probability exponentially close to unit. The above analysis 
implies also that the probability for the eavesdropper to have complete 
information about the key without being detected does not exceed $2^{-2(N-m)}$.

Suppose now that the channel is noisy leading to the errors in the detection 
results obtained by user B, for example, due to the polarization rotation
and let the probability this process is $p<1$ (probability of user A sending
0 and user B detecting 1 and vice versa). As in the previous case, the 
legitimate users accept only the transmissions where no delay
was observed. After completing a series in which $4N$ transmissions
are accepted, users A and B disclose $2N$ transmissions
and estimate the noise level in the channel (probability $p$). Knowledge of 
the error in sending one bit allows in principle (for a sufficiently long
sequence) to choose a suitable {\it classical} block code [19] which actually 
permits to reduce the error in code words to an arbitrarily small value.

For example, user A announces to B through a public channel only the
numbers of transmissions where he sent 1, combining them
into the groups of $2k$ transmissions, and similarly for 0 producing
the blocks of size $2k$ (coded 1 and 0). Then user B applies the majority 
voting to fix the errors in each group (this coding allows to fix $k-1$
errors). The blocks where $k$ errors were detected are rejected (user B
announces the numbers of these blocks through the public channel).
In the groups left after this procedure the error probability does not
exceed $p^k\ll p$. Now the users have the sequences of the groups
(new $\tilde{1}$ and $\tilde{0}$) of length $2\tilde{N}$. 
Then the hashing procedure consisting of $m$ rounds similar to that 
described above is performed resulting in a sequence of length 
$2(\tilde{N}-m)$.  The probability of users A and B having identical
sequences of length $2(\tilde{N}-m)$ under the condition that $m$
rounds of random hashing gave successful parity tests is not less
than $1-2^{-(\tilde{N}-m)}$. The block coding is only required
to increase the surviving probability of the accepted sequence in
the hashing procedure. The hashing procedure can also be applied
to the original (rather than block) sequence. However, in that case
the probability of missing the parity error during hashing will be small
because of the noise. However, if the sequence passes the test, it is 
secure and identical for both users with the above probabilities.
The probability of eavesdropper obtaining complete information about the key
and remaining undetected is well below $2^{-2(\tilde{N}-m)}$ (because
it is sufficient to guess one bit from each code group). This simple coding 
scheme obviously is not the optimal one, but it allows to formulate the
cryptographic protocol in the simplest and most transparent way.

The authors are grateful to Prof. V.L.Golo for the interest to this work
and discussion of the obtained results.

This work was supported by the Russian Foundation for Basic Research
(project No 99-02-18127) 
as well as the project 
``Physical foundations of quantum computer'' 
and Russian Federal Program 
``Advanced technologies and micro- and nanoelectronic devices'' 
(project No 02.04.5.2.40.T.50).

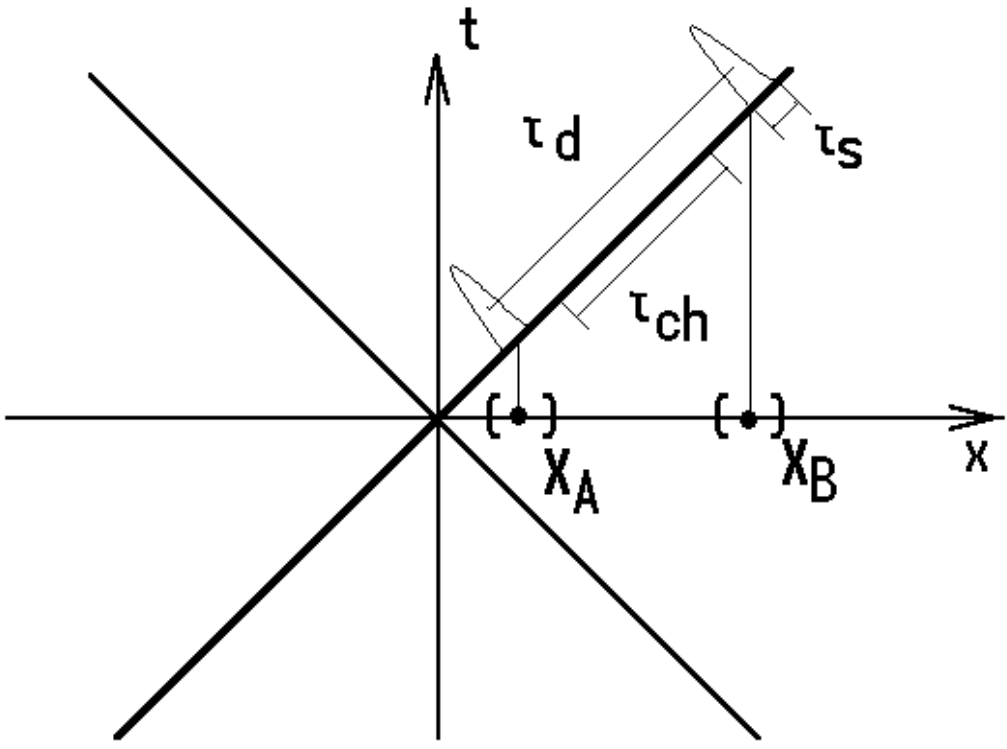
\end{document}